\documentclass[fleqn,usenatbib]{mnras}

\usepackage[T1]{fontenc}
\usepackage{ae,aecompl}

\usepackage{graphicx}	
\usepackage{amsmath}	
\usepackage{amssymb}	
\usepackage{hyperref}
\usepackage{pdflscape}	

\title[Dust-induced pitch-angle reversal]
{Dust-induced reversal of the optical wavelength dependence of spiral-arm pitch angles}

\author[Mart\'{\i}nez-Garc\'ia et al.]{
Eric E. Mart\'inez-Garc\'ia ,$^{1,3}$\thanks{E-mail: ericmartinez@inaoep.mx}
Rosa A. Gonz\'alez-L\'opezlira,$^{2}$,
Iv\^anio Puerari$^{3}$,
and A. I. Gómez-Ruiz.$^{1,3}$
\\
$^{1}$Secretar\'ia de Ciencia, Humanidades, Tecnolog\'ia e Innovaci\'on (SECIHTI), Av. Insurgentes Sur 1582 Colonia Cr\'edito Constructor,\\ 
Benito Ju\'arez, Ciudad de M\'exico, M\'exico, C.P. 03940\\
$^{2}$Instituto de Radioastronom\'ia y Astrof\'isica, UNAM, Campus Morelia, Michoac\'an, M\'exico, C.P. 58089\\
$^{3}$Instituto Nacional de Astrof\'isica, \'Optica y Electr\'onica, Luis E. Erro 1, Tonantzintla, Puebla, M\'exico, C.P. 72840\\
}

\date{Accepted XXX. Received YYY; in original form ZZZ}

\pubyear{2026}

\begin{document}
\label{firstpage}
\pagerange{\pageref{firstpage}-~-\pageref{lastpage}}
\maketitle


\begin{abstract}

The classical density-wave scenario predicts a blue-to-red age--colour gradient across the spiral arms of disc galaxies
as a consequence of shock-induced star formation. However, observations of spiral-arm pitch angles at different wavelengths
sometimes suggest the opposite trend, namely a red-to-blue ``colour jump''.
The origin of this apparent contradiction remains unclear. In this work, we investigate the effects of dust attenuation
on post-shock stellar emission, together with the streamline motions of gas and stellar populations.
We find that dust attenuation in regions with relatively high optical depths, $\hat{\tau}_{V} \gtrsim 0.2$,
can produce an apparent red-to-blue spatial sequence, with respect to the shock location, by reversing the relative positions of the flux maxima
of stellar populations observed in different bands. Moreover,
streamline motions confine post-shock stellar populations near the spiral arms for longer than in a circular-orbit model,
thereby modifying the spatial distribution of the emitted light.
We conclude that the apparent red-to-blue colour jump does not necessarily contradict the classical density-wave scenario.
Instead, it can arise from the same post-shock stellar populations that intrinsically produce a blue-to-red age--colour gradient
if their emission is modified by dust attenuation.

\end{abstract}




\section{Introduction}~\label{sec:intro}

The origin and nature of spiral structure in disc galaxies remain a central problem in astrophysics.
Among the proposed mechanisms, the density wave (DW) theory~\citep{lin64, bert89, shu16} has historically
provided a compelling framework, describing spiral arms as long-lived quasi-stationary patterns that rotate with
a constant angular speed. In this scenario, large-scale shocks compress the interstellar medium, triggering star
formation near the spiral potential minimum~\citep{rob69}.
As a consequence, newly formed stars drift away from their birthplaces,
producing observable azimuthal age (or colour, blue-to-red) gradients across spiral arms.

Early observational efforts to detect these blue-to-red colour gradients yielded mixed results.
However, the development of reddening-insensitive photometric indices, such as the $Q(rJgi)$ index,
enabled more robust tracing of stellar populations~\citep{gg96}.
Using this approach,~\citet{mart09a} analysed a sample of non-barred (SA) and weakly barred (SAB) spiral galaxies,
and found that a significant fraction exhibit blue-to-red colour gradients
consistent with theoretical expectations, supporting the idea that spiral density waves can trigger star formation.
These gradients also provide a means of estimating the spiral pattern speed and locating dynamical resonances,
including corotation, placing key constraints on the physical nature and origin of spiral arms~\citep[e.g.,][]{marc24}.

In parallel, attention has been given to the morphological properties of spiral arms, especially the pitch angle,
defined as the angle between the arm tangent and a circular orbit. The pitch angle is a fundamental observable
that encodes information about galactic dynamics and structure.
Some early studies suggested that spiral-arm pitch angles vary with wavelength~\citep[e.g.,][]{gros98,mart12,mart13}.
Using integral field spectroscopy and multi-band imaging,
\citet{mart14} found that in several galaxies the pitch angle increases towards longer optical wavelengths, indicating a red-to-blue sequence.
Accordingly, within the framework of DW theory, the spatial azimuthal offsets manifest as two seemingly opposite observational signatures:
the classical age (or colour) gradient~\citep[blue-to-red; e.g.,][]{gg96},
and the so-called ``colour jumps''~\citep[red-to-blue; e.g.,][]{yu18}.

\begin{figure} 
\centering
\includegraphics[width=1.0\hsize]{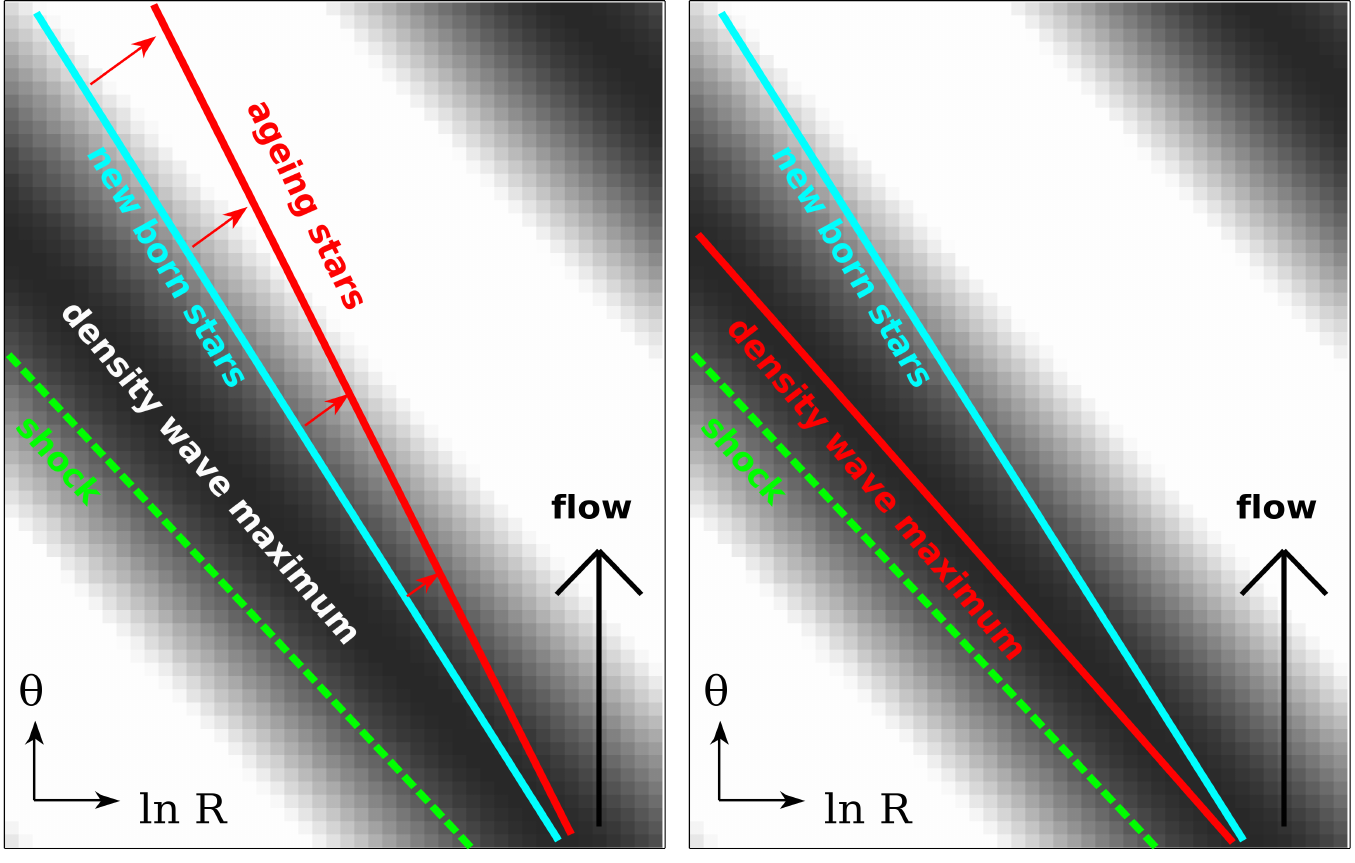}
\caption[fig1.eps]{
Age gradients and colour jumps across trailing spiral arms in the $\ln R$--$\theta$ plane for $R < R_{\mathrm{CR}}$,
where $R_{\mathrm{CR}}$ is the corotation radius.
Left: classic blue-to-red age gradient; the dashed green line, dark grey region, solid blue line,
and solid red line indicate the spiral shock, density wave, newborn stars, and ageing stars, respectively.
Right: red-to-blue colour jump, located upstream of the classic age gradient; the dashed green, solid red,
and solid blue lines indicate the spiral shock, density wave maximum, and newborn stars, respectively.
In this plane, a vertical line corresponds to a circle ($R=\mathrm{constant}$), with
pitch angle $P=0\degr$, while a horizontal line has $P=90\degr$; for the plotted lines,
$P$ increases in the anticlockwise direction.
Adapted from~\citet{mart23}, with permission.
}~\label{fig1}
\end{figure}

One qualitative interpretation of red-to-blue colour jumps comes from assuming that the potential minimum, dominated by old stars,
the shock region, traced by dust lanes, and the sites of recent star formation, traced by young stars, are not co-spatial~\citep{git04}.
Consequently, measurements of the pitch angle at different optical wavelengths, each sensitive to distinct stellar populations,
yield systematically different values (see Figure~\ref{fig1}).

Nevertheless, this qualitative interpretation does not fully account for the observed dichotomy between
blue-to-red colour gradients and red-to-blue colour jumps. In particular, it neglects potentially
important effects, such as dust attenuation, which may bias both the inferred colour gradients and their
physical interpretation. For instance, some studies report smaller absolute pitch angles in the red
than in the blue, a finding that is inconsistent with red-to-blue colour jumps but consistent with
classical blue-to-red gradients~\citep[e.g.,][]{pou16,mil19,abd22}.

Moreover, observations of spiral galaxies in deep fields reveal that both blue-to-red and/or
red-to-blue sequences may be present even at redshifts corresponding to look-back times of several gigayears,
indicating that the physical processes associated with spiral density waves have operated
over a substantial fraction of cosmic history~\citep[][hereafter Paper~I]{mart23}.
These findings suggest that the interplay between spiral structure, star formation,
and stellar population gradients is both fundamental and long-lived in disc galaxies.

Taken together, these studies point towards a coherent but still incomplete picture.
While density wave theory successfully explains many observed features---including colour gradients,
resonance structures, and population offsets---uncertainties remain regarding how observable quantities depend on wavelength.
In particular, reported variations in pitch angle with optical wavelength raise important questions
about the physical interpretation of spiral structure and its dependence on stellar-population tracers.

In this paper, we revisit the wavelength dependence of spiral-arm pitch angles within the context of density wave theory,
building upon previous observational and theoretical work. Our goal is to clarify the physical origin of this dependence
and explore its implications for models of spiral structure and star formation in galaxies.

\section{Analysis}

In this section, we describe the analysis performed to investigate the physical mechanisms that may
explain the aforementioned colour dichotomy, with particular emphasis on the red-to-blue colour jump.
We first simulate the post-shock fluxes of stellar populations of different ages.
We then examine the effects of dust attenuation and streamline motions on the expected azimuthal age--colour gradients,
and finally compute the corresponding spiral-arm pitch angles.

\subsection{Modelling the starlight from a shock-driven sequence of star formation}~\label{starlight_model}

\citet{rob69} described the physical conditions of the post-shock gas across spiral arms,
leading to the prediction of azimuthal age--colour gradients.
To model the post-shock light emitted by stellar populations of different ages,
we first construct the star formation history (SFH) associated with this process.

A simple and flexible description of the post-shock SFH 
can be obtained by using two exponentially declining components.
In this parametrisation, the first exponential represents the underlying,
long-term star formation episode responsible for building most of the stellar mass.
A second exponential component can then be added to describe a more recent burst of star formation~\citep[see e.g.,][]{daC08,boq19},

\begin{equation}~\label{eq:SFH}
  \Psi(t) = \Psi_{\rm old}(t) + \Psi_{\rm burst}(t'),
\end{equation}

\noindent where $\Psi$ is the star formation rate, $t$ is the time since the onset of star formation,
and $t'$ represents time elapsed since the onset of the second star formation burst.
We adopt the following parametrisation:

\begin{eqnarray}
  \Psi_{\rm old}(t) & = & (\tau_{0}^{-1}) \exp(-t/\tau_{0}),\label{eq:SFH_old}\\
  \Psi_{\rm burst}(t')    & = & (k)\exp(-t'/\tau_{1}).\label{eq:SFH_burst}           
\end{eqnarray}

\noindent The parameters $\tau_0$ and $\tau_1$ are the e--folding timescales associated with the
old stellar component and the recent star formation episode, respectively.
The parameter $k$ controls the relative normalisation of the second exponential component,
defined as $k\propto\frac{f}{1-f}$~\citep{boq19}, where

\begin{equation}
f=\frac{\int_{0}^{t'}\Psi_{\rm burst}(t'){\rm d}t'}
{\int_{0}^{t}\Psi(t) {\rm d}t},
\end{equation}

\noindent i.e., the fraction of stellar mass formed in the second burst
relative to the total stellar mass ever formed.

\begin{figure} 
\centering
\includegraphics[width=1.0\hsize]{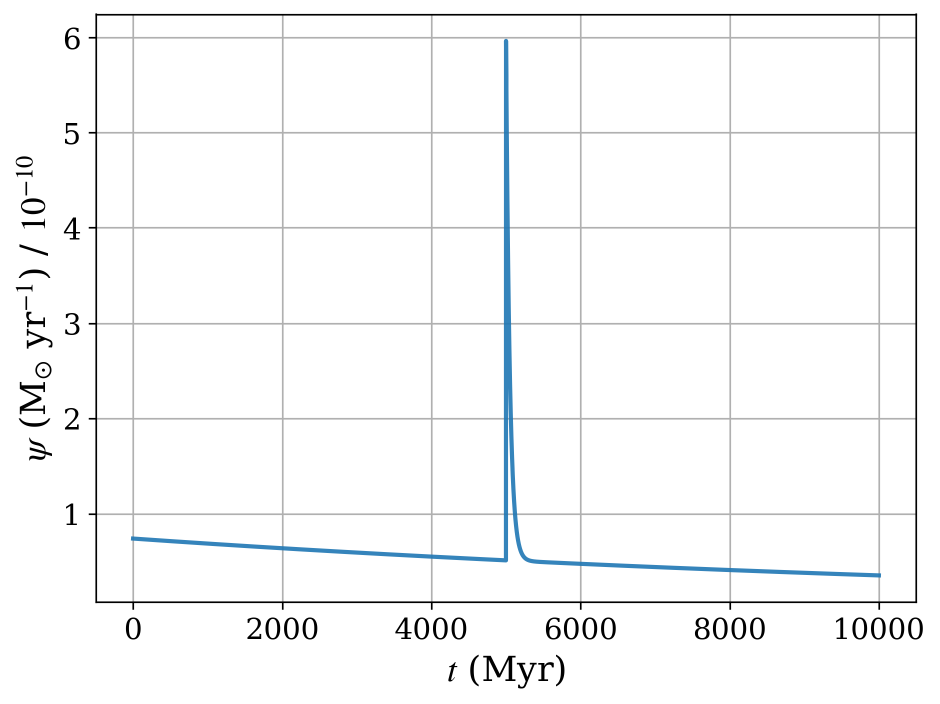}
\caption[fig2.eps]{
Star formation history (SFH) used to model the spiral-shock event at $t=5\times10^3$ Myr.
Here, $\Psi(t)$ denotes the star formation rate, and $t$ is measured from the onset of star
formation in the underlying old stellar population.
A recent burst of star formation is superimposed on this old stellar component.
}
\label{fig2}
\end{figure}

To model the age--colour gradient, we adopt the SFH parameters listed in Table~\ref{tbl-1},
where T$_{\rm form}$ denotes the present-day age of the old stellar component,
and B$_{\rm form}$ corresponds to the present-day age of the stars formed during the recent burst.
The resulting SFH is shown in Fig.~\ref{fig2}. For the recent burst, we adopt $\tau_1=50$ Myr;
this implies that, at $t'\simeq35$ Myr, the star formation rate $\Psi_{\rm burst}(t')$ has decreased to
approximately $50\%$ of its initial value.\footnote{
\citet{mart09a} modelled the gradient using an old stellar population with an age of $5\times 10^3$ Myr
and a recent burst duration of 20 Myr, with young stars contributing between 1 and 5 per cent of the stellar mass.
These parameters are broadly consistent with those adopted here.}

\begin{figure} 
\centering
\includegraphics[width=1.0\hsize]{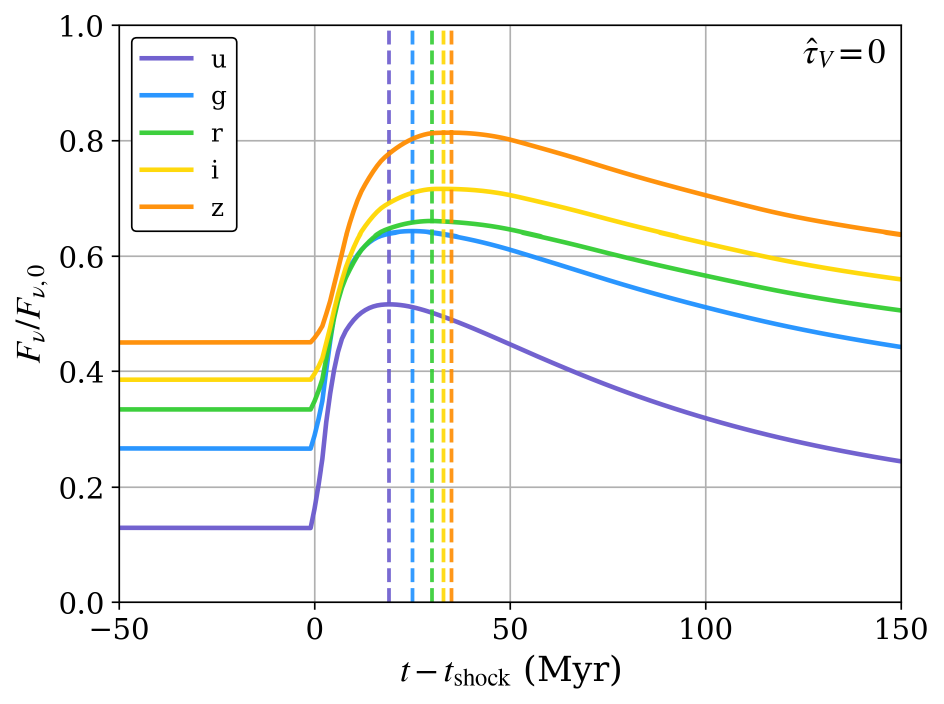}
\caption[fig3.eps]{
Theoretical stellar emission in the $ugriz$ optical bands (solid lines) resulting from the SFH shown in Figure~\ref{fig2}.
For each band, the time of maximum relative flux is indicated by a dashed vertical line.
The flux densities are normalised by $F_{\nu,0}=5\times10^{4}\,\mathrm{mJy}$.
These models do not include dust attenuation ($\hat{\tau}_{V} = 0$).
The figure shows only the results of the SPS models and does not account for stellar dynamics near the spiral arms.
}~\label{fig3}
\end{figure}

Based on this SFH, we model the stellar emission using the stellar population synthesis (SPS)
models of~\citet{bc03}. We adopt the BASEL stellar library, a~\citet{chab03} initial mass
function (IMF), and solar-metallicity templates ($Z_{\sun}\simeq~0.02$). We then simulate the
stellar flux in the Sloan Digital Sky Survey (SDSS) $ugriz$ optical bands~\citep{fuk96,doi10}, see Table~\ref{tbl-2}.
All calculations are performed using the~\textsc{Code Investigating Galaxy Emission}~\citep[\textsc{CIGALE};][]{nol09,boq19}.\footnote{
Identical results were obtained using the \textsc{GALAXEV} code of~\citet{bc03}.}
In this case, we assume $\hat{\tau}_{V} \simeq 0$; models with $\hat{\tau}_{V} > 0$ are discussed
in~\S~\ref{dust_model}. The results of this exercise are shown in Fig.~\ref{fig3},
where we highlight the time of maximum relative flux in each band, motivated by the fact that
observational pitch-angle measurements are intensity-weighted~\citep{yu18}.
A blue-to-red temporal sequence is recovered, with $(t-t_{\rm shock}) = 19, 25, 30, 33$, and $35~{\rm Myr}$
for the $u$, $g$, $r$, $i$, and $z$ bands, respectively, as expected for a classical age--colour gradient.
Once the velocity vector is specified, this temporal sequence can be directly mapped into the
corresponding spatial sequence across the spiral arm.
 
\begin{table}
	\centering
	\caption{SFH parameters used to model the age--colour gradient.}
	\label{tbl-1}
	\begin{tabular}{ccccc} 
		\hline
		$\tau_{0}$ (Myr) & $\tau_{1}$ (Myr) & $f$ & T$_{\rm form}$ (Myr) & B$_{\rm form}$ (Myr)\\ 
		\hline
		$13.5\times10^3$ & 50 & 0.05 & $1\times10^4$ & $5\times10^3$\\
		\hline
	\end{tabular}
\end{table}

\begin{table}
	\centering
	\caption{Filter Characteristics}
	\label{tbl-2}
	\begin{tabular}{cccccc} 
		\hline
		Band & $u$ & $g$ & $r$ & $i$ & $z$\\ 
		\hline
		$\lambda_{\mathrm{eff}}$ (\AA) & 3608 & 4672 & 6141 & 7458 & 8923\\
		\hline
	\end{tabular}
\end{table}

\subsection{Dust attenuation}~\label{dust_model}

We investigate the effect of dust attenuation on the observed age--colour gradients at optical wavelengths.
For this purpose, we adopt the two-component
dust attenuation model of~\citet{cha00}.
This model accounts for the age-dependent attenuation of stellar
populations. Observational studies indicate that giant molecular clouds are generally short-lived structures,
with characteristic lifetimes of approximately 10-30 Myr~\citep[e.g.,][]{che20}.
After the emergence of massive stars, stellar feedback typically disperses the parent
molecular cloud within a few Myr.
Accordingly, it is assumed that newly formed stars remain embedded within their natal birth clouds for
a characteristic timescale of $t_{\mathrm{BC}}\simeq10^{7}\,\mathrm{yr}$.
Their radiation is therefore attenuated by dust in both the birth clouds
(BCs) and the diffuse ambient interstellar medium (ISM). Older stars, whose
birth clouds have dispersed, are attenuated only by the diffuse ISM. The
effective optical depth can be written as
\begin{equation}
\hat{\tau}_{\lambda}(t^{*}) =
\begin{cases}
\hat{\tau}^{\mathrm{BC}}_{\lambda}
+\hat{\tau}^{\mathrm{ISM}}_{\lambda},
& t^{*} \leq t_{\mathrm{BC}},\\[3pt]
\hat{\tau}^{\mathrm{ISM}}_{\lambda},
& t^{*} > t_{\mathrm{BC}},
\end{cases}
\end{equation}
where $t^{*}$ is the stellar age. Both components are commonly described by
power-law attenuation curves $\hat{\tau}_{\lambda}\propto\lambda^{\delta}$,
with typical values of $\delta_{\rm{ISM}}=-0.7$, and $\delta_{\rm{BC}}=-1.3$~\citep[see, e.g.,][]{daC08}.

The dust attenuation is specified by two parameters: the total effective attenuation optical
depth in the $V$ band, $\hat{\tau}_{V}$, and the fraction of this optical depth contributed
by the ambient interstellar medium (ISM), defined as

\begin{equation}
  \mu=\hat{\tau }_V^{\,\mathrm{ISM}}/(\hat{\tau}_V^{\,\mathrm{BC}}+\hat{\tau}_V^{\,\mathrm{ISM}}),
\end{equation}

\noindent where $\hat{\tau }_V^{\,\mathrm{BC}}$ and $\hat{\tau }_V^{\,\mathrm{ISM}}$
denote the contributions from stellar birth clouds and the ambient ISM, respectively.

\begin{figure} 
\centering
\includegraphics[width=1.0\hsize]{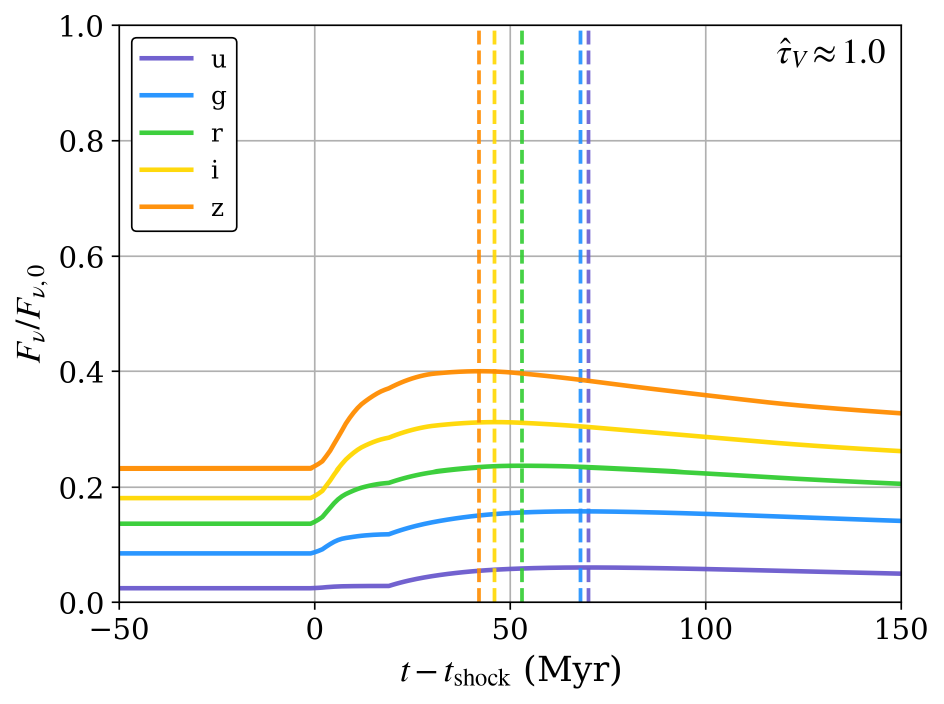}
\caption[fig4.eps]{
Theoretical stellar emission in the optical $ugriz$ bands (solid lines) resulting from the
SFH shown in Figure~\ref{fig2}, after attenuation by dust.
The attenuation was computed using the two-component dust absorption model of~\citet{cha00}.
The adopted dust parameters are $\hat{\tau}_{V} \approx 1.0$, $\mu = 0.5$, $\delta_{\rm{ISM}}=-0.7$,
and $\delta_{\rm{BC}}=-3.0$. For each band, the time of maximum relative flux is
indicated by a dashed vertical line. Flux densities are normalised by $F_{\nu,0}=5\times10^{4}\,\mathrm{mJy}$.
As in Figure~\ref{fig3}, this plot does not include the stellar dynamics near the spiral arms, but only the results of the SPS models.
}~\label{fig4}
\end{figure}

\begin{figure*}
\centering
\includegraphics[width=1.0\hsize]{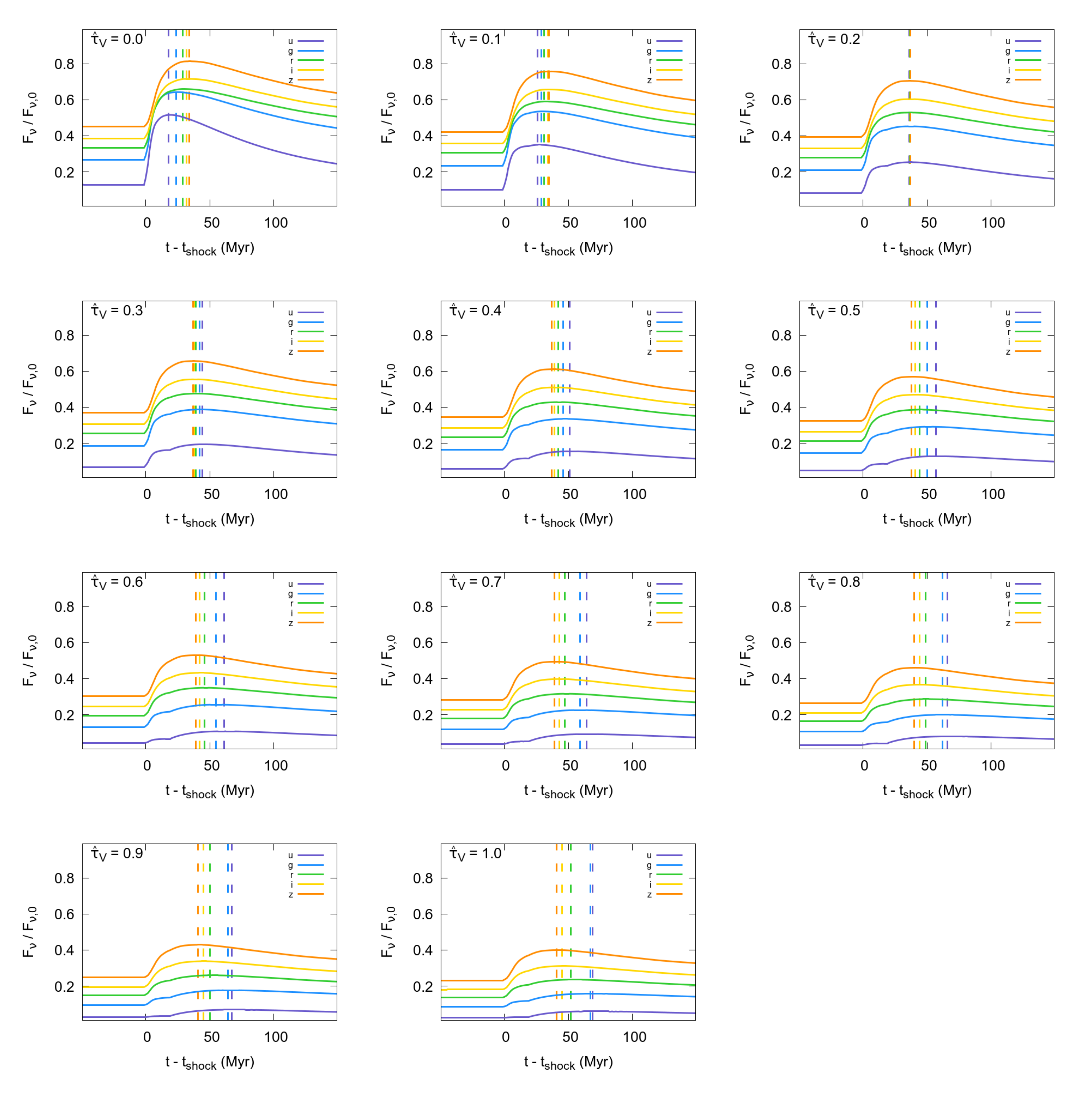}
\caption[fig5.eps]{Theoretical stellar emission in the optical bands for different dust optical depths,
ranging from $\hat{\tau}_{V}=0$ to $\hat{\tau}_{V}=1.0$, resulting from the SFH shown in Figure~\ref{fig2}.
For each model, the time of maximum relative flux in each band is indicated by dashed vertical lines.
The sequence of flux maxima changes from blue-to-red, in dust-free or weakly attenuated models, to
red-to-blue in more strongly attenuated models ($\hat{\tau}_{V} \gtrsim 0.2$).
All flux densities are normalised by $F_{\nu,0}=5\times10^{4}\,\mathrm{mJy}$.
As in Figure~\ref{fig3}, this plot does not include the stellar dynamics near the spiral arms, but only the results of the SPS models.
}~\label{fig5}
\end{figure*}

We combine the two-component dust attenuation prescription with the post-shock stellar emission model described 
in \S~\ref{starlight_model}. The result is shown in Fig.~\ref{fig4},
adopting $\hat{\tau}_{V} \approx 1.0$, $\mu = 0.5$, $\delta_{\rm{ISM}}=-0.7$, and $\delta_{\rm{BC}}=-3.0$.
The steeper wavelength dependence assumed for $\hat{ \tau}_\lambda^{\, \mathrm{BC}}$ is motivated by the higher
ambient densities surrounding stellar birth clouds in the shock and post-shock regions~\citep[e.g.,][]{shu73},
relative to typical unperturbed environments.
In this case, the relative flux reaches its maximum at $(t-t_{\rm shock}) = 70, 68, 53, 46$, and $42~{\rm Myr}$
in the $u$, $g$, $r$, $i$, and $z$ bands, respectively. The ordering of these peak times is therefore
reversed relative to that obtained for the $\hat{\tau}_{V}= 0$ case (see Figure~\ref{fig3}).
In Figure~\ref{fig5}, we show the stellar emission predicted for different values of $\hat{\tau}_{V}$,
ranging from $\hat{\tau}_{V}=0$ to $\hat{\tau}_{V}=1.0$. For $\hat{\tau}_{V} \gtrsim 0.2$, the ordering of the flux maxima changes from
a blue-to-red to a red-to-blue sequence.

\subsection{Streamline motions}

In the~\citet{rob69} spiral-shock framework, gas streamlines denote the flow trajectories of interstellar gas in the frame
corotating with the spiral density wave. These streamlines trace the gas motion as it crosses the spiral-arm shock,
where the gas is compressed, decelerated, and deflected, and subsequently follows a post-shock orbit downstream of the arm.
This behaviour was studied by~\citet{mart09b}, to assess its effect on azimuthal age--colour gradients,
and on the estimation of pattern speeds and resonance locations in spiral galaxies.

In the density-wave scenario, the galactic disc is subject to a fixed spiral potential rotating at a constant global pattern speed.
Under this assumption, \citet[][see their fig.~3]{dob10} found that post-shock gas and stars remain close to,
and flow approximately along, the spiral arm for $\sim50$~Myr, before moving into the interarm region on a timescale of $\sim100$~Myr.

\begin{figure} 
\centering
\includegraphics[width=1.0\hsize]{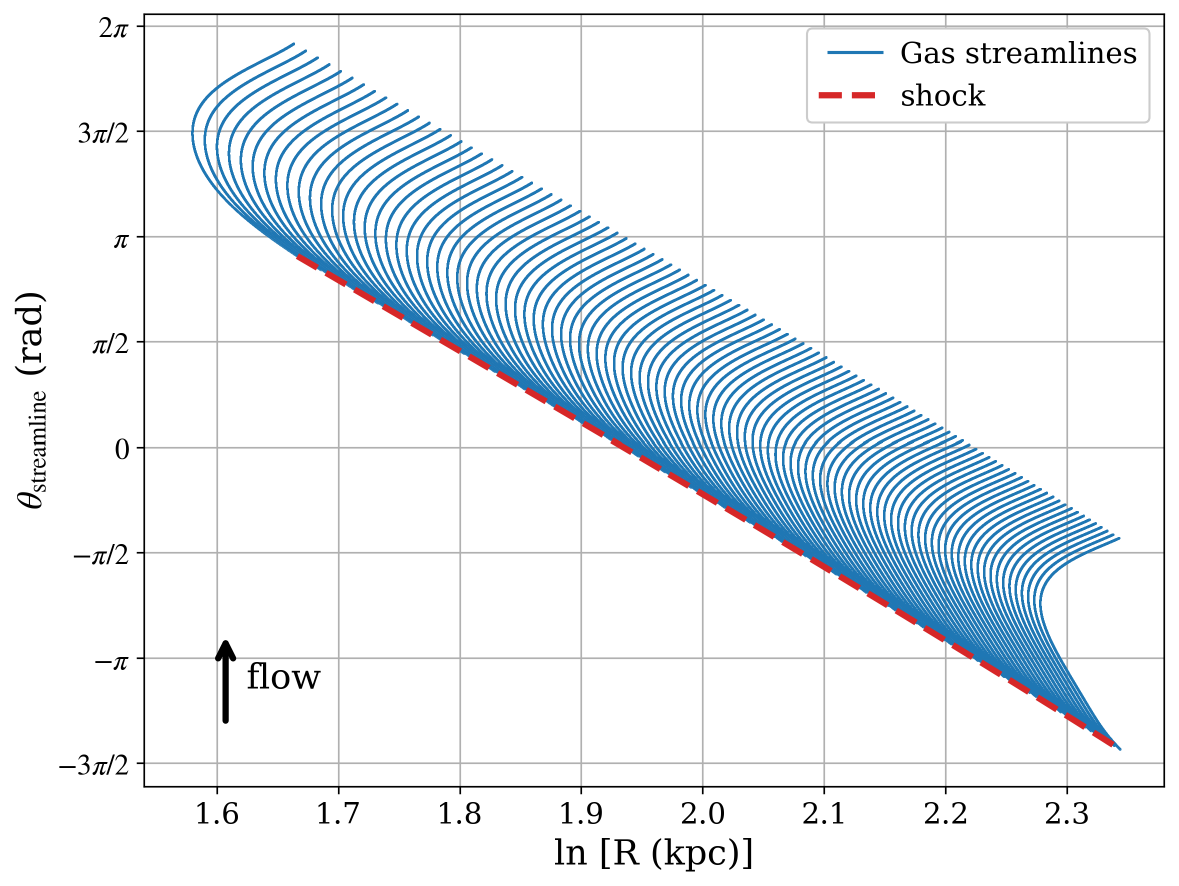}
\caption[fig6.eps]{
Gas streamlines after crossing the spiral shock, for $R < R_{\mathrm{CR}}$.
Solid (blue) lines show the post-shock gas trajectories,
which are tangent to the local gas velocity field in the reference
frame rotating with the spiral pattern.
The dashed red line marks the location of the spiral shock,
and the vertical black arrow indicates the sense of rotation/flow.
}~\label{fig6}
\end{figure}

Typical gas streamlines are shown in Figure~\ref{fig6}.
As expected, after crossing the shock, the gas streamlines are deflected predominantly along the spiral arm,
rather than across it, towards smaller galactocentric radii.
To compute these streamlines, we use the semi-analytical solutions for spiral shocks
adopted by~\citet{mart09b}, who followed the earlier work of~\citet{rob69},
\citet{shu72},~\citet{shu73}, and~\citet{git04}. The adopted parameters are:
number of spiral arms, $m=2$; spiral-arm pitch angle, $\sin i = 0.1$;
angular speed of the spiral pattern, $\Omega_{\rm p}=13$ km s$^{-1}$ kpc$^{-1}$;
effective sound speed, $a=8$ km s$^{-1}$; ratio of the perturbed spiral-field amplitude
to the axisymmetric field, $F=0.05$, assumed to be constant; and circular rotation
velocity of the material, $v_{\rm rot}=220$ km s$^{-1}$. Similar gas streamlines are
obtained in magnetohydrodynamic (MHD) numerical simulations using a two-arm spiral
potential with a pitch angle of $15.5\degr$ and $\Omega_{\rm p}=20$ km s$^{-1}$ kpc$^{-1}$~\citep[see][]{mart09b}.
In the present investigation, we assume that newly formed post-shock stars initially follow trajectories
that closely trace the corresponding gas streamlines.

\begin{figure} 
\centering
\includegraphics[width=1.0\hsize]{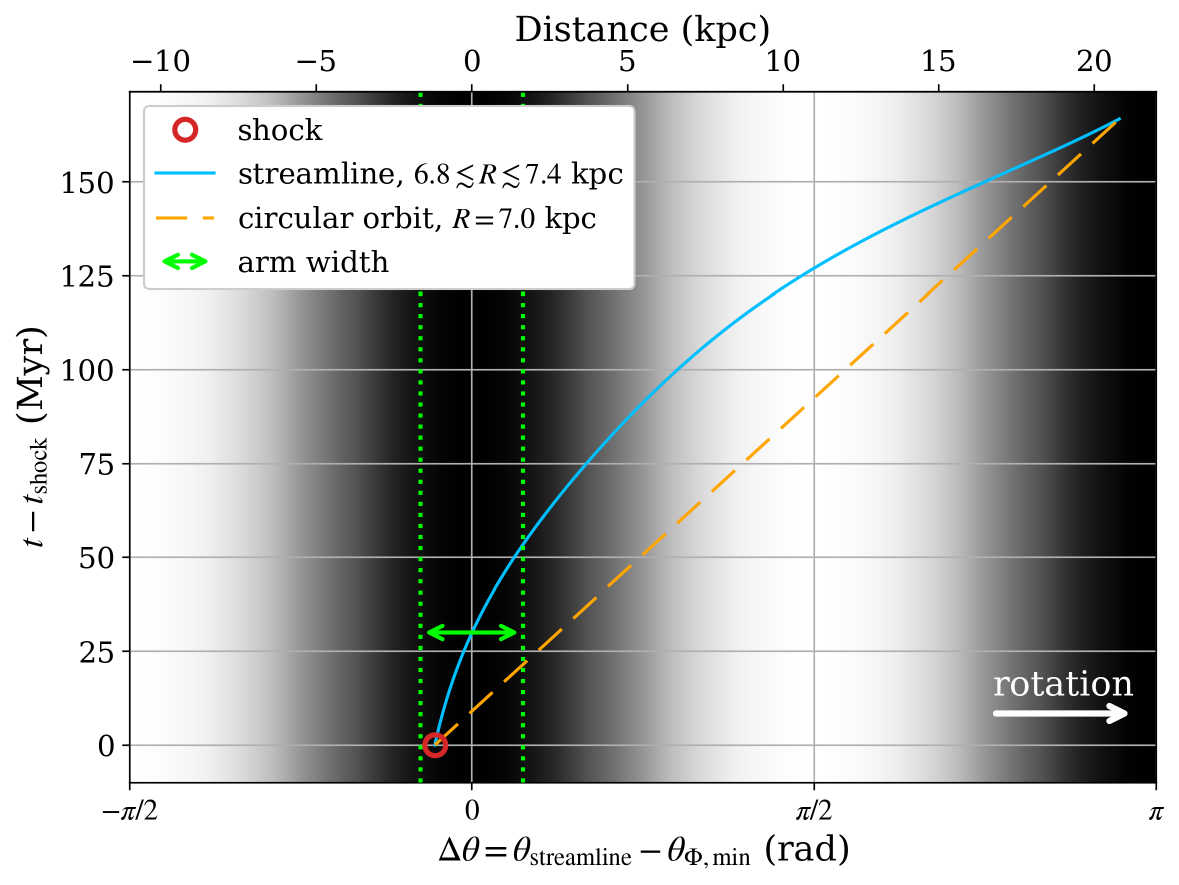}
\caption[fig7.eps]{
Time elapsed since crossing the spiral shock as a function of the angular offset,
$\Delta\theta=\theta_{\rm streamline}-\theta_{\Phi,\min}$, within $6.8\lesssim R \lesssim 7.4$~kpc.
The solid cyan line traces the gas streamline, and the red open circle marks the location of the
spiral shock at $(t-t_{\rm shock})=0$.
The orange long-dashed line shows the expected circular orbit at $R=7$~kpc relative to the spiral pattern.
The $m=2$ spiral density wave is indicated by the grey-scale shaded region.
The green short-dashed lines delimit an arm width of $\sim3.3$~kpc.
The white arrow indicates the direction of rotation.
}~\label{fig7}
\end{figure}

In Figure~\ref{fig7}, we show the time elapsed since passage through the spiral shock as a function of the angular offset,
$\Delta\theta=\theta_{\rm streamline}-\theta_{\Phi,\min}$, where $\theta_{\Phi,\min}$ denotes the potential minimum,
for a representative streamline within $6.8\lesssim R \lesssim 7.4$~kpc.
The motion relative to the spiral pattern is compared with that expected for a circular orbit at $R=7$~kpc.
The grey-scale background represents an $m=2$ spiral density wave proportional to $\cos(2\Delta\theta)$,
with maximum density (black) at $\Delta\theta=0$ and minimum density (white) at $\Delta\theta=\pi/2$.
We also indicate, with the green vertical short-dashed lines,
the mean spiral-arm width measured in galaxy samples, $w=3.3\pm1.2~\mathrm{kpc}$~\citep{sav20}.
The streamline motion retains post-shock stellar populations within the spiral-arm width for up to $(t-t_{\rm shock})\sim53\pm7$~Myr,
compared with only $\sim21\pm4$~Myr for circular orbits, thereby modifying the spatial distribution of the emitted light.
Moreover, because the stellar populations move predominantly along the spiral arm rather than across it during the
initial post-shock phase, the optical depth, $\hat{\tau}_{V}$, experienced by these populations remains nearly constant.
This contrasts with the circular-orbit case, in which stars move directly across the spiral structure
and therefore experience more rapid variations in optical depth as they pass from the arm to the interarm regions,
for which $\hat{\tau}^{\rm{arm}}_{V}/\hat{\tau}^{\rm{interarm}}_{V}\approx 2$--$3$~\citep[e.g.,][]{dom00,gon03}.

\subsection{Pitch angles of the post-shock isochrones}

\begin{figure} 
\centering
\includegraphics[width=1.0\hsize]{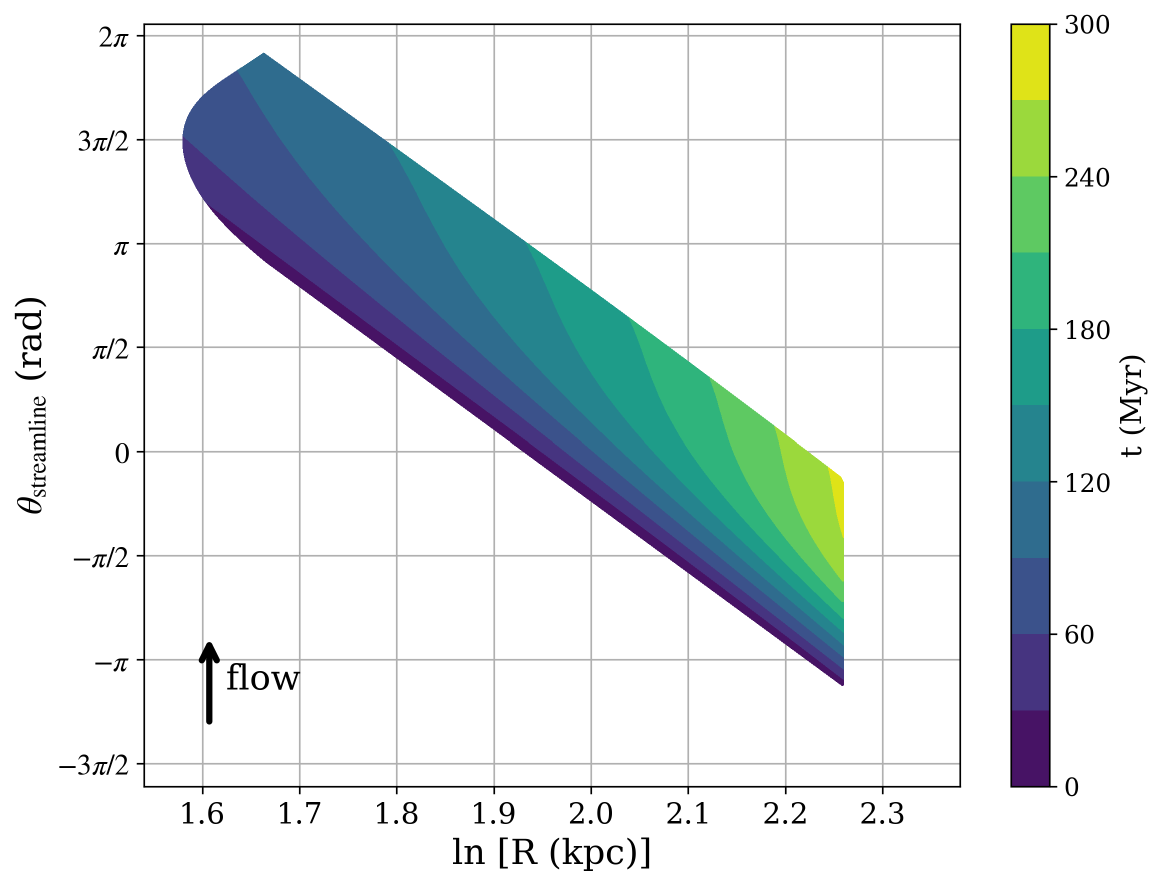}
\caption[fig8.eps]{
Isochrone curves connecting points on different gas streamlines (see Figure~\ref{fig6}),
with the same elapsed orbital time since crossing the spiral shock at $t=0$.
The vertical black arrow indicates the sense of rotation/flow.
}~\label{fig8}
\end{figure}

To mimic the observational measurement of spiral-arm pitch angles, we proceed as follows.
We define post-shock isochrones as loci of equal travel time along the gas flow,
measured from the shock-crossing point on each streamline.
These curves connect positions on different gas streamlines that correspond to the same
elapsed time after passage through the spiral-arm shock, as shown in Figure~\ref{fig8}.
We then identify the post-shock isochrones associated with the times of maximum relative flux
in each band (see Figures~\ref{fig3},~\ref{fig4}, and~\ref{fig5}).
The pitch angles are then computed from these post-shock isochrones as described below.

We compute the pitch angle of the post-shock isochrones, corresponding to each of the $ugriz$ bands,
by fitting the arm locus in the $(\ln R,\theta)$ plane.\footnote{In real observations,
spiral-arm pitch angles are commonly measured from
deprojected galaxy images by transforming the light distribution to $(\ln R,\theta)$ coordinates and applying a
two-dimensional Fourier transform, where the dominant Fourier frequency is converted to pitch angle assuming
logarithmic spiral arms~\citep[e.g.,][]{pue92,sar94,mart14}.}
For each isochrone a linear relation of the form
\begin{equation}
    \theta = a \ln R + b
\end{equation}
was fitted. Since
\begin{equation}
    a = \frac{d\theta}{d\ln R},
\end{equation}
the logarithmic-spiral pitch angle is given by
\begin{equation}
    P = \arctan\left(\frac{d\ln R}{d\theta}\right)
      = \arctan\left(\frac{1}{a}\right).
\end{equation}

In order to remove the dependence on the winding direction of
the spiral arm, we quote the absolute value of the pitch angle,
    $|P|_{\rm deg}
    =
    \left|
    \frac{180}{\pi}
    \arctan\left(\frac{1}{a}\right)
    \right|$.
This procedure was applied to the shock locus, and
the post-shock isochrones in the $ugriz$ bands.

The uncertainty in $P$ was estimated by propagating the uncertainty
in the fitted slope.  The linear fit returns the covariance matrix of
the fitted parameters,
\begin{equation}
    \mathbf{C} =
    \begin{pmatrix}
        \sigma_a^2 & {\rm cov}_{ab} \\
        {\rm cov}_{ba} & \sigma_b^2
    \end{pmatrix},
\end{equation}
so that the $1\sigma$ uncertainty in the slope is
\begin{equation}
    \sigma_a = \sqrt{C_{00}} .
\end{equation}

Because the pitch angle depends only on the slope $a$, the intercept $b$
uncertainty does not contribute directly. Standard error
propagation gives
\begin{equation}
    \sigma_P =
    \left|
    \frac{{\rm d}P}{{\rm d}a}
    \right| \sigma_a =
    \frac{\sigma_a}{1+a^2},    
\end{equation}
where $\sigma_P$ is in radians.

\begin{figure} 
\centering
\includegraphics[width=1.0\hsize]{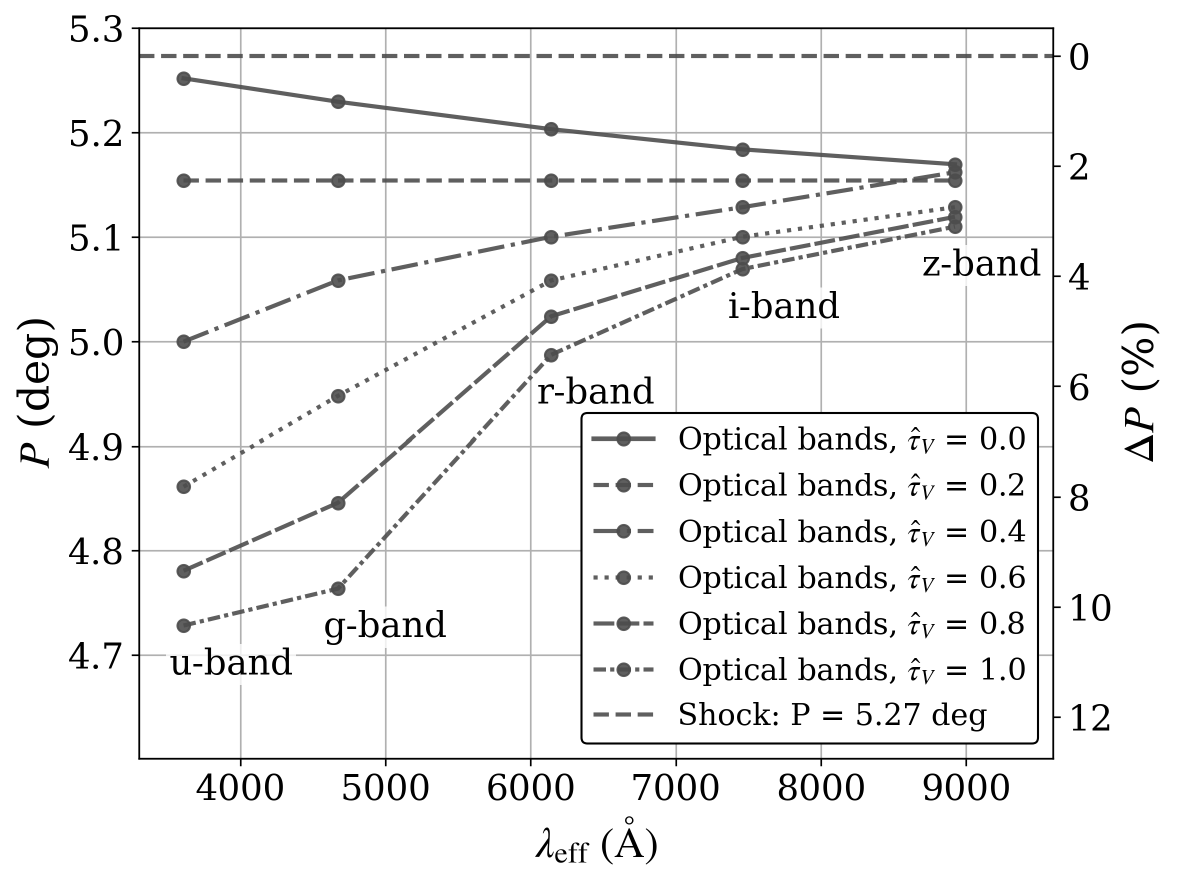}
\caption[fig9.eps]{
Resulting pitch angles ($P$) measured at the time of maximum relative flux in each optical band
for models with different dust optical depths, $\hat{\tau}_{V}\simeq 0$--$1.0$
(see Figure~\ref{fig5}). The pitch angles are derived from the post-shock gas/star orbits,
following the streamline and isochrone geometry shown in Figures~\ref{fig6} and~\ref{fig8}.
The right-hand $y$-axis gives the relative pitch-angle difference, $\Delta P = |P_{\rm shock}-P|/P_{\rm shock}\times 100$, in per cent.
The uncertainties in $P$ are smaller than $0.5\%$.
}~\label{fig9}
\end{figure}

Using the corresponding $ugriz$ isochrones, we derive the pitch angles shown in Figure~\ref{fig9}.
For $\hat{\tau}_{V} \lesssim 0.2$, the pitch angle decreases with increasing wavelength,
as expected for a classical blue-to-red age--colour gradient, whereas for $\hat{\tau}_{V} \gtrsim 0.2$,
the pitch angle increases with increasing wavelength.
We therefore show that an apparent red-to-blue colour jump can arise from an underlying
blue-to-red age--colour gradient through the combined effects of dust attenuation and streamline motions.

\section{Discussion}

In this section, we discuss the physical interpretation of the main results presented in this work.
We first consider the stellar populations responsible for the two apparently opposite age--colour gradients,
arguing that both arise from the same underlying stellar populations and that their observed differences are primarily
driven by dust attenuation. We then investigate whether the observed colour dichotomy could be attributed to the presence of
leading spiral arms and assess this possibility in the context of the strong observational preference for trailing spiral structure.
Finally, we explore the non-monotonic, or ``wavy'', dependence of spiral-arm pitch angle on wavelength
(see Paper~I) and investigate whether variations in the properties of the post-shock dust environment can account for this behaviour.

\subsection{Colour reversal without an age reversal}

Paper~I interpreted the red-to-blue sequence as a configuration in which old stars,
tracing the density wave, lie upstream of newly born stars, thereby producing an apparent ``colour jump''.
In this scenario, the post-shock ageing stellar population would have to contribute a relatively weak
radiation field compared with that of the old stars associated with the density wave.
Thus, the blue-to-red and red-to-blue sequences would be produced by an old and an ageing stellar population, respectively,
but by the same population of newly born stars (see Figure~\ref{fig1}).

As shown in Figure~\ref{fig4}, the relative stellar flux in the reddest band, $z$,
is approximately a factor of two higher at its maximum than at $(t-t_{\rm shock})=0$, i.e. immediately before the shock.
In addition, observed arm-to-interarm contrasts are generally modest in the old stellar component,
with near-infrared or stellar-mass contrasts typically amounting to only a few tenths of a per cent~\citep{bit17,que24}.
Therefore, the stellar flux from the newly formed stars and their ageing
descendants remains higher than that from the old stars tracing the density wave.

Taken together, these considerations suggest that the apparent colour dichotomy does not
require a reversal of the underlying stellar-age sequence. Instead, it depends on whether
the stellar populations are embedded in a dusty environment. Dust attenuation can reverse the
relative ordering of the observed flux maxima in different bands,
producing an apparent red-to-blue sequence downstream of the shock. The underlying stellar
populations nevertheless continue to age downstream, as expected.
Thus, despite the reversal in the observed colour sequence, the stellar populations dominating the bluer bands,
e.g. $u$ and $g$, at $(t-t_{\rm shock})\sim50$--$70$ Myr are located further downstream and are therefore older
than those dominating the redder bands, e.g. $i$ and $z$, at $(t-t_{\rm shock})\sim40$ Myr.

\subsection{The leading-arm interpretation}

By measuring the pitch angles of galaxies over the redshift range
$z=0.2$--$1.2$, Paper~I (their Section~4.1) identified some
systems exhibiting blue-to-red colour jumps rather than the more commonly
observed red-to-blue sequence. They considered whether this behaviour could
be explained by the presence of leading spiral arms. However, our results
disfavour this interpretation.

Observational determinations of the near sides of inclined discs, combined
with spectroscopic velocity fields, have long indicated that spiral arms are
overwhelmingly trailing rather than leading
\citep[e.g.][]{deV58,pash82,sha85}. This conclusion has been supported by
more recent studies. For example, \citet{iye19} found that all galaxies in
their sample were consistent with trailing arms under the standard
dust-lane interpretation of the near side. Confirmed or plausible
leading-arm systems, such as NGC~4622, therefore appear to be rare
exceptions~\citep{but03,vai08}. This observational evidence is also
consistent with the results of~\citet{mei24}, who found that the physical
conditions in galactic discs preferentially amplify trailing spiral
structure.

We therefore favour an alternative explanation in which dust attenuation
determines the apparent direction of the colour jump. Regions with
relatively high optical depths, $\hat{\tau}_{V} \gtrsim 0.2$, may exhibit an
apparent red-to-blue colour sequence, whereas regions with
$\hat{\tau}_{V} \lesssim 0.2$ may retain the intrinsic blue-to-red sequence.
In this scenario, the observed red-to-blue sequence does not indicate a
distinct stellar population. Instead, it arises from the same underlying
stellar populations responsible for the intrinsic blue-to-red gradient,
but viewed through regions containing greater amounts of dust.

\subsection{Wavy trends in the pitch angle--wavelength relation}

The relation between the absolute value of the spiral-arm pitch angle, $|P|$,
and the rest-frame optical wavelength, $\lambda_{\rm rest}$, is not always adequately
described by a simple linear trend. Although linear fits provide a useful first-order
characterisation of the overall behaviour, the full multi-wavelength measurements
reveal a more complex, non-monotonic, or ``wavy'', pattern~(see figs~15 and 16 of Paper~I).
This effect is relevant for pitch-angle measurements in both nearby and high-redshift galaxies,
and may reflect variations in the physical conditions of the post-shock dust environment.

\begin{figure} 
\centering
\includegraphics[width=1.0\hsize]{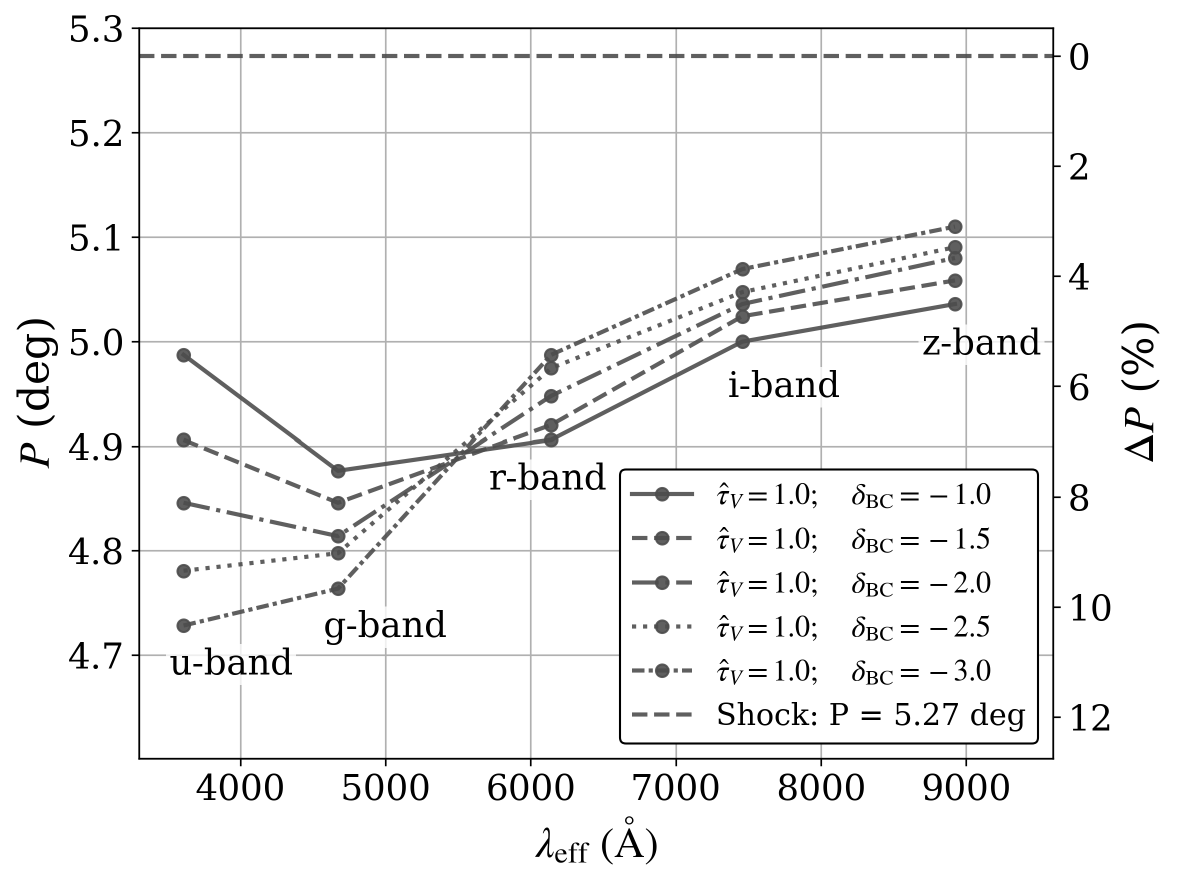}
\caption[fig10.eps]{
Resulting pitch angles ($P$) measured at the time of maximum relative flux in each optical band
for models with $\hat{\tau}_{V}=1.0$ and different values of $\delta_{\rm{BC}}$ (see \S~\ref{dust_model}).
The pitch-angle computations take into account the post-shock gas/stellar orbits.
The right-hand $y$-axis shows the relative pitch-angle difference, $\Delta P$, in per cent.
The uncertainties in $P$ are smaller than $0.5\%$.
}~\label{fig10}
\end{figure}

\begin{figure} 
\centering
\includegraphics[width=1.0\hsize]{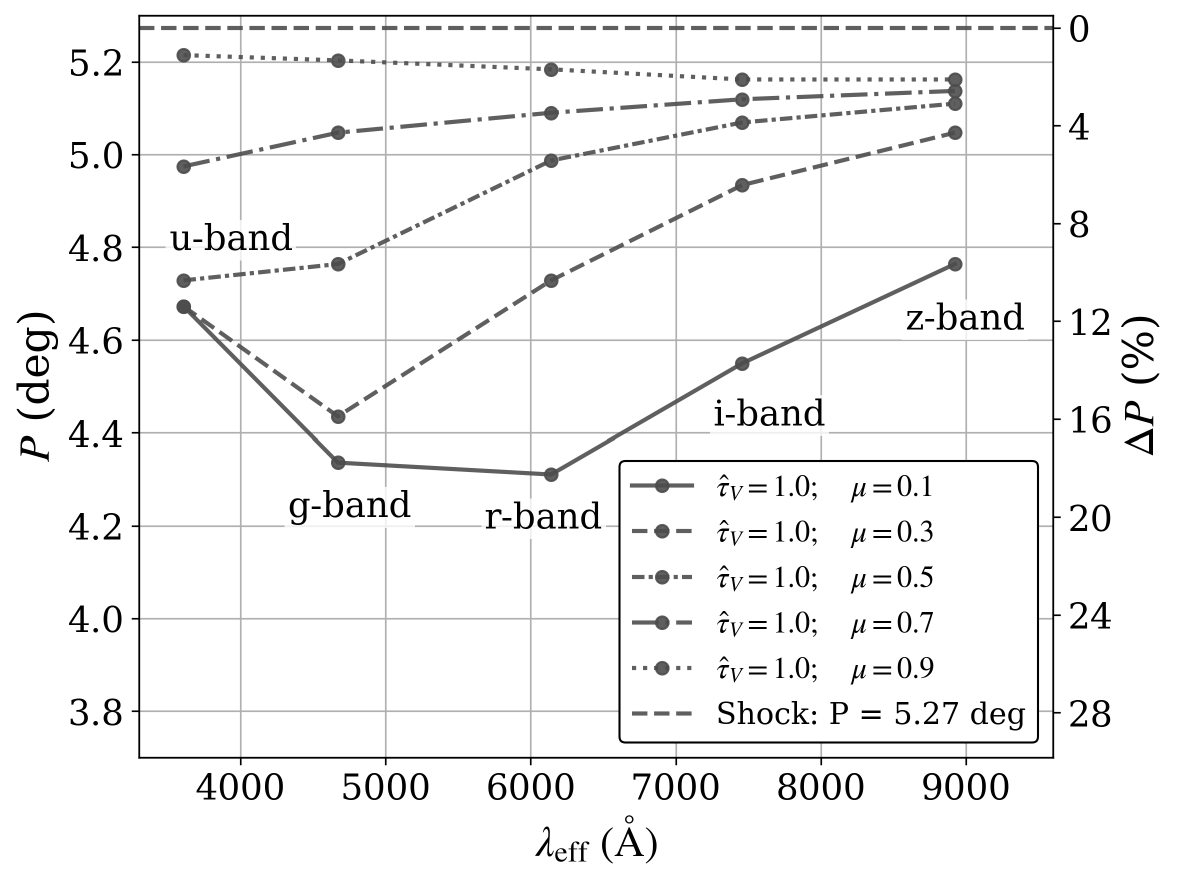}
\caption[fig11.eps]{
Similar to Figure~\ref{fig10}, but for different values of $\mu$ (see \S~\ref{dust_model}).
The right-hand $y$-axis shows $\Delta P$ in per cent.
The uncertainties in $P$ are smaller than $0.5\%$.
}~\label{fig11}
\end{figure}

To investigate this effect, Figures~\ref{fig10} and~\ref{fig11} show how the pitch-angle measurements
change when varying the dust-model parameters. In particular, we explore values of $\delta_{\rm BC}$
from $-1$ to $-3$ and $\mu$ from 0.1 to 0.9 (see \S~\ref{dust_model}).
The results show a non-monotonic behaviour similar to the ``wavy'' patterns reported in~Paper~I.
This suggests that the observed ``wavy'' trends may be explained, at least in part, by spatial variations
in the physical conditions of post-shock dust across spiral arms.

\section{Conclusions}

In the spiral density-wave scenario for disc galaxies, spiral shocks trigger star formation and produce an
age--colour gradient across the spiral arms. We have investigated the physical origin of the observed
colour-gradient dichotomy: while some galaxies exhibit a blue-to-red signature,
corresponding to the classical gradient, others show an apparent red-to-blue sequence, or colour jump,
in which the absolute value of the pitch angle increases towards longer optical wavelengths.

We first analysed the effect of dust attenuation using the two-component model of~\citet{cha00}.
For optical depths $\hat{\tau}_{V} \gtrsim 0.2$, we found that dust attenuation selectively shifts
the location of the maximum relative flux from newly formed stars farther from the shock for
the stellar populations observed in the bluer bands.

We then analysed the effect of streamline motions on the location of the maximum relative flux from
newly formed stars relative to the shock, assuming that these stars follow trajectories similar to
those of the post-shock gas. We found that streamline motions modify the positions of the stellar
populations relative to the spiral arms, causing them to remain in the vicinity of the arms for
longer than they would on circular orbits.

The net effect of dust and streamline motions is an increase in the $P$ values towards longer optical wavelengths,
producing an apparent red-to-blue sequence of stellar populations in the azimuthal direction.
We therefore conclude that \textit{red-to-blue colour jumps can be understood as classical blue-to-red
age--colour gradients observed within dusty environments}.

In addition to optical studies of stellar populations, complementary 
observations at other wavelengths or star formation tracers and/or numerical studies
~\citep[e.g.][]{egu09,hon15,wal16,ho17,res22,val22,res23,gar24,sun24,bob25,chu25,deb25,rosh25,sun25,ard26}
may provide further constraints on the origin and evolution of spiral structure (see also Appendix~\ref{appendix_A}).

An important related question is whether spiral patterns are long-lived density waves that persist for
several Gyr~\citep{lin64,lin66}, or transient features that survive for only a few orbital
periods~\citep[e.g.][]{sel11,ros12,sel14,sel22}. The presence of a classical bulge may promote
the formation and/or longevity of two-armed spiral patterns~\citep[e.g.,][]{smi24}. In addition,
some studies have suggested that spiral structure in the local Universe may result
from the combined action of multiple mechanisms~\citep{che24a,che24b,que25}.
Nevertheless, grand-design spiral galaxies have been identified across a broad redshift range,
$1\lesssim z\lesssim 5$~\citep{daw03,law12,elm14,yua17,hod19,tsu21,mar22,fud22,wu23,kuh24,kuh26,kal25,jai25,dero25,xia25}.

In this context, one may ask what minimum lifetime a spiral mode must have to generate observable colour
gradients across spiral arms, such as those reported in statistically significant samples of disc galaxies.
Although we do not address this question quantitatively in the present work, it provides important motivation
for our analysis by placing spiral-arm colour gradients within the broader context of the origin and evolution of spiral structure.

\section*{Acknowledgements}

We thank the referee for their valuable comments, which have helped to improve the manuscript.
EMG acknowledges support through the ``Investigadoras e Investigadores por México'' (formerly Cátedras CONACYT) programme,
which is now supported by the Secretaría de Ciencia, Humanidades, Tecnología e Innovación (SECIHTI), in Mexico.
RAGL acknowledges the financial support of DGAPA, UNAM, project IN106124.

\section*{Data availability}

There are no new data associated with this article.








\appendix
\section{Spiral density waves at millimetre wavelengths}~\label{appendix_A}

An important implication of the age--colour-gradient model presented in this work is that
it can be used to predict the pitch angles of spiral galaxies at wavelengths beyond the optical regime.
In particular, observations at millimetre wavelengths (1000 $\micron$) trace emission from cold dust.
If spiral density waves trigger star formation across the spiral arms, the cold dust
surrounding the star-forming regions may absorb the UV and optical radiation associated
with the stellar age gradient and re-emit this energy at millimetre wavelengths.
Consequently, the spatial offset encoded by the age gradient may produce a characteristic spiral-arm pitch angle in the millimetre emission.

To investigate this scenario, we used~\textsc{CIGALE}~\citep{nol09,boq19}.
For the stellar emission and dust attenuation, we adopt the same parameters as those
described in~\S~\ref{starlight_model} and~\S~\ref{dust_model}, respectively.
For the dust emission, we adopt the two-parameter model of~\citet{dal14}.
In this model, the first parameter, $\alpha_{\rm SF}$, controls the infrared--submillimetre--radio spectral energy distributions of star-forming galaxies.
It is defined through the relation $dM_{\rm{d}}\propto~U^{-\alpha_{\rm SF}}dU$,
where $M_{\rm{d}}$ is the dust mass heated by a radiation field of intensity $U$. Larger values of $\alpha_{\rm SF}$
correspond to lower characteristic dust-grain temperatures. We adopt $\alpha_{\rm SF}=2.0$, which is approximately
the midpoint of the allowed range, $0.06\leq\alpha_{\rm SF}\leq4.0$. The second parameter in the~\citet{dal14} model
specifies the fractional contribution of an active galactic nucleus (AGN) to the total dust emission; we set this contribution to zero.

\begin{figure}
\centering
\includegraphics[width=1.0\hsize]{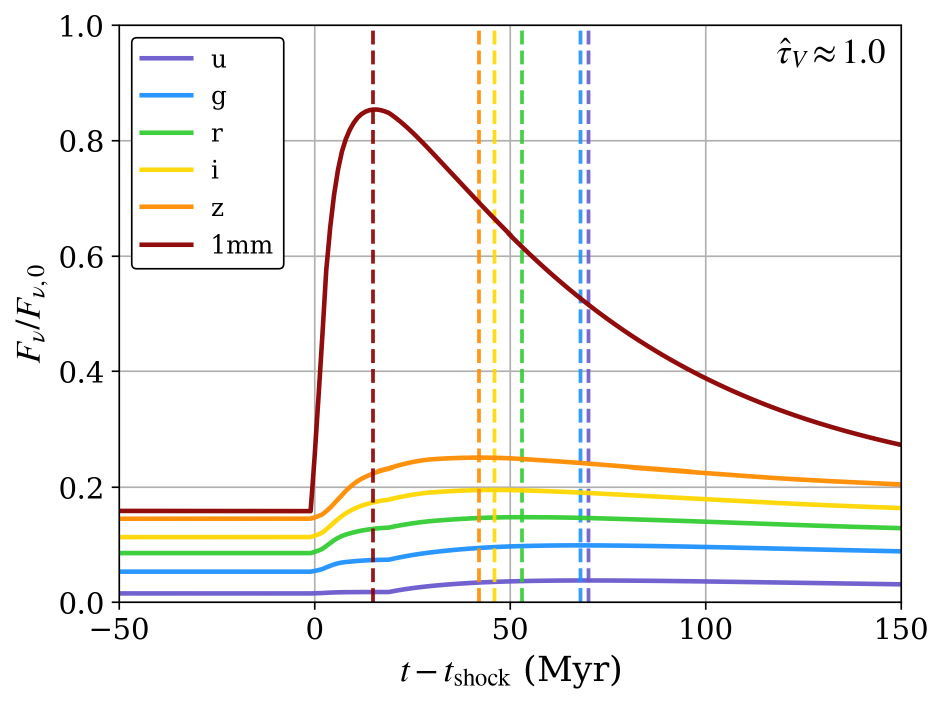}
\caption{
Similar to Figure~\ref{fig4}, but including the predicted emission in the MUSCAT 1-mm band.
For each band, the time at which the relative flux reaches its maximum is indicated by a dashed vertical line.
The flux densities are normalised by $F_{\nu,0}=8\times10^{4}~\mathrm{mJy}$.
}
\label{figA1}
\end{figure}

Figure~\ref{figA1} shows the generalised predicted emission at 1.1 mm.
This wavelength is widely accessible to current and upcoming continuum instruments,
such as the MUSCAT~\citep{bri18} and TolTEC~\citep{bry18} cameras on the
50-m Large Millimeter Telescope Alfonso Serrano~\citep[LMT;][]{hug20}, among other millimetre facilities. 
This result suggests that, when red-to-blue colour jumps are observed,
the spiral arms traced at millimetre wavelengths should exhibit pitch angles larger than
those measured in the optical, but smaller than those of the shock locus.
Comparing the spiral-arm pitch angles measured at 1~mm with those obtained at other wavelengths for a statistically
significant sample of galaxies could provide an important observational test of the age--colour-gradient model
and help constrain the physical origin and evolution of spiral density waves in galaxies.

\bsp	
\label{lastpage}
\end{document}